\documentstyle[12pt]{article}

\def\labelmark{}
\def\void{}
\newenvironment{formula}[1]{\def\labelname{#1}
\ifx\void\labelname\def\junk{\begin{displaymath}}
\else\def\junk{\begin{equation}\label{\labelname}}\fi\junk}%
{\ifx\void\labelname\def\junk{\end{displaymath}}
\else\def\junk{\end{equation}}\fi\junk\labelmark\def\labelname{}}
{\ifx\void\labelname\def\junk{\end{array}\end{displaymath}}
\else\def\junk{\end{array}\right.\end{equation}}
\fi\junk\labelmark\def\labelname{}\def\junk{}
}
\newenvironment{formulae}[1]{\def\labelname{#1}
\ifx\void\labelname\def\junk{\begin{displaymath}}
\else\def\junk{\begin{eqnarray}\label{\labelname}}\fi\junk}%
{\ifx\void\labelname\def\junk{\end{displaymath}}
\else\def\junk{\end{eqnarray}}\fi\junk\labelmark\def\labelname{}}

\newcommand{\ba}{\begin{array}}
\newcommand{\ea}{\end{array}}
\newcommand{\beq}{\begin{formula}}
\newcommand{\eeq}{\end{formula}}
\newcommand{\beqa}{\begin{formulae}}
\newcommand{\eeqa}{\end{formulae}}

\def\BC{\bb C}
\def\_\BC{\bbi C}

\newcommand{\De}{\Delta}
\newcommand{\ep}{\epsilon}

\newcommand{\pa}{\partial}
\newcommand{\al}{\alpha}

\begin{document}
\begin{titlepage}
\setcounter{page}{1}
\renewcommand{\thefootnote}{\fnsymbol{footnote}}

\begin{flushright}
IC/2000/168\\
hep-th/0011055
\end{flushright}

\vspace{13mm}
\begin{center}
{\Large Graded $q$-pseudo-differential Operators
and Supersymmetric Algebras} 
\vspace{14mm}

{\large Ahmed Jellal$^{a}$ 
\footnote{E-mail: jellal@ictp.trieste.it -- jellal@youpy.co.uk }}
\,{and}\,
{\large El Hassan El Kinani$^{b}$ 
\footnote{E-mail: el$_{-}$kinani@fste.ac.ma}}\\
\vspace{5mm}
{\em $^{a}$ High Energy Physics Section\\
 the Abdus Salam International Centre for Theoretical Physics\\
 Strada Costiera 11, 34100 Trieste, Italy} \\
\vspace{5mm}
{\em $^{b}$ Mathematical Physics Group, Dept. of Mathematics\\ 
Faculty of Sciences and Technics, University Molay Ismail \\
P.O.B. 509, Errachidia, Morocco}
\end{center}

\vspace{5mm}
\begin{abstract}
We give a supersymmetric generalization of the sine algebra and 
the quantum algebra $U_{t}(sl(2))$. Making use of the 
$q$-pseudo-differential operators graded with a fermionic algebra,
we obtain a supersymmetric extension of sine algebra. With this 
scheme we also get a quantum superalgebra $U_{t}(sl(2/1)$.
\end{abstract}

\vspace{8mm}
\vfill
\begin{flushleft}
\end{flushleft}

\end{titlepage}
\newpage
%%%%%%%%%%%%%%%%%%%%%%%%%%%%%%%%%%%%%%%%%%%%%%%%%%%%%%%%%%%%%%%%%%%%%%%
\setcounter{footnote}{0}
\renewcommand{\theequation}{\thesection.\arabic{equation}}
\renewcommand{\theequation}{\arabic{equation}}
%--------------------------------------------------------------------
\section{Introduction}\label{sec1}
\setcounter{section}{1}
\setcounter{equation}{0}
\indent
%-------------------------------------------------------------------
%%%%%%%%%%%%%%%%%%%%%%%%%%%%%%%%%%%%%%%%%%%%%%%%%%%%%%%%%%%%%%%%%%%%
%                                                                  %
%                              Text                                %
\indent                                                            %
%%%%%%%%%%%%%%%%%%%%%%%%%%%%%%%%%%%%%%%%%%%%%%%%%%%%%%%%%%%%%%%%%%%%
One of the most important infinite-dimensional Lie algebra is the one
generated by the so-called pseudo-differential operators {\cite{1}}.
This can be viewed as a generalization of the Virasoro algebra
and of the Lie algebra of differential operators. Recently, 
the supersymmetric algebras have been applied to the study of some physical 
problems. For example, the supersymmetric sine algebra is used to 
investigate the properties of Bloch electron in a constant uniform 
magnetic field {\cite{2}}. Moreover, the quantum superalgebras are 
applied to solve some problems for instance related to superconductivity 
{\cite{3} and the quantum Hall effect {\cite{4}}. These results 
lead us to think to the present work.

In this paper, we will present an approach to obtain a 
realization of certain supersymmetric algebras. More precisely, 
we will propose a graded $q$-pseudo-differential operator 
realization of the supersymmetric extension of the sine algebra
and the quantum superalgebra $U_{t}(sl(2/1))$. 

This paper is organized as follows: In section $2$ we review some 
basic notions related to $q$-pseudo-differential operators and also to
the realization of sine algebra and $U_{t}(sl(2))$ in this framework. 
We propose a realization of the supersymmetric extension of the last 
algebras in section $3$. We give a conclusion in the final section.
                                                              
\section{Prelimenaries}
Before going on, we would like to give a short review concerning 
some basic notions, which will be useful in the next section. This 
concerns the $q$-pseudo-differential operators, sine algebra and 
$U_t(sl(2))$.

\subsection{$q$-pseudo-differential operators}
We start first by defining the so-called $q$-derivation. For this matter,
let $q$ be a complex number different from $0$ and $1$. By definition,
the $q$-derivation or more generally $\alpha$-derivation is given by
\beq{.d}
d_{\al}(fg)=\al(f)d_{\al}(g)+d_{\al}(f)g,
\eeq
where $f, g$ $\in C[x,x^{-1}]$ are the ring of polynomials in an 
indeterminant $x$ and its inverse $x^{-1}$. In eq.(1), $\al$ is a 
linear mapping. An example of $\al$-derivation is given by Jackson's 
$q$-differential operator $\pa_{q}$, such as {\cite{5}}
\beq{p}
\pa_{q}(f)={f(qx)-f(x)\over (q-1)x},
\eeq
which leads to the following form for eq.$(1)$
\beq{.d}
\pa_{q}(fg)=\eta_{q}(f)\pa_{q}(g)+\pa_{q}(f)g,
\eeq
where the shift operator $\eta_{q}$ is 
\beq{.1}
\eta_{q} (f(x))=f(qx).
\eeq

Now let us introduce the $q$-pseudo-differential operators 
algebra $q-\psi DO$. The latter is characterized by the 
relation {\cite{5}} 
\beq{.q}
q-\psi DO =P(x,\partial_{q})=
\sum_{i=-\infty}^{N}P_{i}(x)\partial_{q}^{i},\qquad P_{i}(x) \in C[x,x^{-1}].
\eeq
Consequently, the algebra $q-\psi DO$ is generated by
$x$, $x^{-1}$, $\pa_{q}$, $\pa_{q}^{-1}$ with the 
relation
\beq{.p}
\pa_{q}x-qx\pa_{q}=1.
\eeq
Note that the family $\{x^{i}\pa_{q}^{j}\}_{i,j\in Z}$ forms a basis 
of $q-\psi DO$. Then, the algebra $q-\psi DO$ can be viewed as a Lie 
algebra, which can be defined by the commutation relation
\beq{.P} 
[P,Q]=P \circ Q-Q \circ P,
\eeq
where the multiplication law "$\circ$" is 
\begin{equation}
\begin{array}{cc}
\pa_{q}\circ f= \eta_{q}(f)\pa_{q}+\pa_{q}f, \\
\pa_{q}^{-1}\circ f=\sum_{k \ge 0}(-1)^{k}q^{-k(k+1)/2}
(\eta_{q}^{-k-1}(\pa_{q}^{k}f)\pa_{q}^{-k-1}.
\end{array}
\end{equation}
The last equation are obtained by using the following relation
\beq{.p} 
\pa_{q}^{-1}\circ\pa_{q}\circ f=\pa_{q}\circ\pa_{q}^{-1}\circ f=f.
\eeq 
Note that eq.$(8)$ can be unified as follows
\begin{equation}
\partial_{q}^{n}\circ f=\sum_{{k \ge 0}}{n\choose k}_{q}(\eta_{q}^{n-k}
(\partial_{q}^{k}f)\partial_{q}^{n-k},
\end{equation}
for all $n$. In the last equation, the $q$-binomials take the form
\beq{.n}
{n\choose k}_{q}=\frac{(n)_{q}(n-1)_{q}....(n-k+1)_{q}}
{(1)_{q}(1)_{q}...(k)_{q}},
\eeq
and the $q$-numbers are given by
\beq{.n}
(n)_{q}=\frac{q^{n}-1}{q-1},
\eeq
where the convention 
\beq{.n}
{n\choose 0}_{q}=1,
\eeq
is taken. We also add that the residue of the symbol 
$P(x,\pa_{q})$ can be written as 
\beq{.R}
{\textbf{Res}}(\sum_{i=-\infty}^{N}P_{i}(x)\partial_{q}^{i}\})=P_{-1}(x),
\eeq
and its {\textbf{Tr}}-functional is 
\begin{equation}
{\textbf{Tr}}(\sum_{i=-\infty}^{N}P_{i}(x)\partial_{q}^{i})=
\int_{s^{1}}P_{-1}(x)dx.
\end{equation}
Considering a subfamily of $q-\psi DO$ as 
\begin{equation}
q-S \psi DO =\{P(x,\partial_{q})=\sum_{i=-\infty}^{N}P_{i}(x).
(\eta_{q})^{i}/Tr(P)=0\}.
\end{equation}
From eq.$(6)$, we obtain the relation
\beq{.e} 
\eta_{q}x=qx\eta_{q}.
\eeq
Therefore, $\eta_{q}$ and $x$ generate a non-commutative algebra,
which is homomorphic with the Manin's plane "quantum plane" {\cite{6}}.

\subsection{Sine algebra}
In this subsection, we review the realization of the sine algebra
and the quantum algebra $U_{t}(sl(2))$. To do this, let us consider
a subfamily of $q-S\psi DO$ generated by $J_{\bf m}$. This
can be constructed as follows {\cite{7}}
\begin{equation}
J_{\bf m}=q^{-m_{1}.m_{2}/2}\eta _{q}^{m_{1}}x^{m_{2}},
\end{equation}
with ${\bf m}=(m_{1},m_{2})$. Calculating the commutation relation, 
it is found
\begin{equation}
[J_{\bf m},J_{\bf n}]=(q^{({\bf m}\times{\bf n})/2}-
q^{-({\bf m}\times{\bf n})/2})J_{{\bf m}+{\bf n}},
\end{equation}
where ${\bf m}\times {\bf n}=m_{1}n_{2}-m_{2}n_{1}$. It is interesting 
to note that when $q$ is a $F$-th root of unity, i.e. 
$q=\exp(\frac{4\pi i}{F})$, the last equation
takes the following form
\begin{equation}
[J_{\bf m},J_{\bf n}]=2i \sin({\frac{2\pi}{F}}
{\bf{m} \times \bf{n}})J_{\bf{m+n}},
\end{equation}
which generates the sine algebra or Fairlie-Fletcher-Zachos 
(FFZ) {\cite{8}} algebra. This is exactly the Moyal 
bracket quantization of the area-preserving diffeomorphisms 
or symplectomorphisms algebra on $2-d$ torus. It should be 
mentioned that the deformation here (eq.$(20)$) is the Moyal 
quantization which is strongly different from the Drinfel'd 
and Jimbo one {\cite{9}}, where the Hopf structure plays a 
crucial role. 
 
Now let us give a construction of the quantum algebra $U_{t}(sl(2))$
in this scheme. Before going on, we recall that this
algebra is defined by the commutation relations~{\cite{9}}  
\begin{equation}
\begin{array}{cccc}
[X^{+},X^{-}]=\frac{t^{2h}-t^{-2h}}{t-t^{-1}},\\

[\ h,X^{\pm}]=\pm X^{\pm},
\end{array}
\end{equation}
where $t$ is the deformed parameter, $t\ne 0,1$. 
In the limit where $t\to 1$, the above equations reduce 
to ones defining the Lie algebra $sl(2)$. 

\noindent The generators of $U_{t}(sl(2))$ can be embedded as 
follows {\cite{7}}
\begin{equation}
\begin{array}{cccc}
X^{+}=\frac{J_{\bf m}-J_ {\bf n}}{t-t^{-1}},\qquad
X^{-}=\frac{J_{-\bf m}-J_{-\bf n}}{t-t^{-1}},\\
t^{+2h}=J_{\bf{m-n}},\qquad
t^{-2h}=J_{\bf{n-m}}.
\end{array}
\end{equation}
they satisfy the commutation relations given by eq.$(21)$ if
$t$ is given by 
\beq{.t}
t=q^{{\bf m}\times{\bf n}/2}.
\eeq
This concludes a short review of the $q$-pseudo-differential operators 
and its related algebras. Now let us address our main goal, which
will be expanded in the next sections.
 
\section{Supersymmetric extension}
In this section we start our generalization of the above results. Otherwise, 
we are looking for a supersymmetric extension of the sine algebra and 
the quantum algebra $U_{t}(sl(2))$. To do this let us begin by the realization
of the supersymmetric sine algebra. This task is the subject of the following
subsection. 

\subsection{Supersymmetric sine algebra}
Our goal here is to extend the sine algebra to supersymmetric case. 
For this matter, let us introduce a fermionic algebra. 
We begin by considering the following matrices {\cite{10}}
\begin{equation}
f=\pmatrix{0&1\cr
        0&0\cr},\qquad
f^+=\pmatrix{0&0\cr
        1&0\cr},
\end{equation}
it is easy to show that the generators $f$ and $f^+$ expanding a
fermionic algebra. The latter is characterized by the relations
\begin{equation}
ff^++f^+f=1,\qquad f^2=0=(f^+)^2.
\end{equation}

To obtain the supersymmetric extension of the operators 
$J_{\bf m}$, we can  proceed as follows
\begin{equation}
\begin {array}{cccc}
T_{\bf m}=J_{\bf m}\otimes(ff^++f^+f),\\
S_{\bf m}=J_{\bf m}\otimes(ff^+-f^+f),\\
\end{array}
\end{equation}
which define the so-called graded $q$-pseudo-differential operators.
These operators can be written as 
\begin{equation}
\begin{array}{cccc}
T_{\bf m}=J_{\bf m}\otimes
\pmatrix{1&0\cr
        0&1\cr},\qquad
S_{\bf m}=J_{\bf m}\otimes
\pmatrix{1&0\cr
        0&-1\cr}.\\
\end{array}
\end{equation}
Using the last equation, we prove that the operators $T_{\bf m}$ and 
$S_{\bf m}$ satisfy the following relations
\begin{equation}
\begin{array}{ccc}
[T_{\bf m},T_{\bf n}]_{-}=(q^{({\bf m}\times {\bf n})/2}-
q^{-( {\bf m}\times{\bf n})/2})T_{\bf m+n},\\

[T_{\bf m},S_{\bf n}]_{-}=(q^{({\bf m}\times {\bf n})/2}-
q^{-({\bf m}\times {\bf n})/2})S_{\bf m+n},\\

[S_{\bf m},S_{\bf n}]_{+}=(q^{({\bf m}\times {\bf n})/2}+
q^{-({\bf m} \times {\bf n})/2})T_{\bf m+n}.\\
\end {array}
\end{equation}
Let us remember that our goal here is to obtain a generalization of
the sine algebra given by eq.$(20)$. For this matter, we consider a
$q$ $F$th root of unity. In this case, we show that the relations 
are verified
\begin{equation}
\begin{array}{ccc}
[T_{\bf m},T_{\bf n}]_{-}=2i\sin({2\pi\over F}
({\bf m} \times {\bf n})T_{\bf m+n}, \\

[T_{\bf m},S_{\bf n}]_{-}=2i\sin({2\pi\over F}
({\bf m} \times {\bf n})S_{\bf m+n}, \\

[S_{\bf m},S_{\bf n}]_{+}=2\cos({2\pi\over F}
({\bf m} \times {\bf n})T_{\bf m+n}. \\
\end{array}
\end{equation}
This set of equations generate exactly the supersymmetric sine algebra 
{\cite{11}}. In the next sextion we show how to obtain $U_t(sl(2/1))$ .

\subsection{Quantum superalgebra $U_t(sl(2/1))$}
The task of the present section is to realize the quantum superalgebra
$U_t(sl(2/1))$ based on the defined supersymmetry operators Eq.(26). We begin
by recalling that the quantum superalgebra $U_t(sl(2/1))$ can be viewed
as a $t$-deformation of the classical Lie superalgebra $sl(2/1)$ through the
$t$-deformed relations, between a set of generators denoted by
$e_{1}$, $e_{2}$, $f_{1}$, $f_{2}$, $k_{1}=t^{h_{1}}$, $k_{1}^{-1}=
t^{-h_{1}}$, $k_{2}=t^{h_{2}}$ and  $k_{2}^{-1}=t^{-h_{2}}$, such as 
{\cite{12}}
\begin{equation}
\begin{array}{c}
k_{1}k_{2}=k_{2}k_{1},\qquad
k_{i}e_{j}k_{i}^{-1}=t^{a_{ij}}e_{j},\qquad
k_{i}f_{j}k_{i}^{-1}=t^{-a_{ij}}f_{j},\\
e_{1}f_{1}-f_{1}e_{1}={k_{1}^2-k_{1}^{-2}\over t-t^{-1}},\qquad
e_{2}f_{2}+f_{2}e_{2}={k_{2}^2-k_{2}^{-2}\over t-t^{-1}},\\
e_{1}f_{2}-f_{2}e_{1}=0,\qquad
e_{2}f_{1}-f_{1}e_{2}=0,\qquad
e_{2}^{2}=0=f_{2}^{2},\\
e_{1}^{2}e_{2}-(t+t^{-1})e_{1}e_{2}e_{1} +e_{2}e_{1}^{2}=0,\qquad
f_{1}^{2}f_{2}-(t+t^{-1})f_{1}f_{2}f_{1} +f_{2}f_{1}^{2}=0.\\
\end{array}
\end{equation}
The last two relations are called the Serre relations. The
matrix $(a_{ij})$ is the Cartan one of $sl(2/1)$, i.e.
\begin{equation}
(a_{ij})=\pmatrix{2&-1\cr
        -1&0\cr}.
\end{equation}
$U_t(sl(2/1))$ is a quasi-triangular Hopf superalgebra endowed 
with the ${\bf Z}_2$-graded Hopf algebra structure
\begin{equation}
\begin{array}{c}
\De(k_i)=k_i\otimes k_i,\qquad \De(e_i)=e_i\otimes{\bf 1}+
e_i\otimes k_i,\qquad \De(f_i)=e_i\otimes k_i^{-1}+{\bf 1}\otimes f_i,\\
\ep(k_i)={\bf 1},\qquad \ep(e_i)=\ep(f_i)=0,\\
S(e_i)=-k_i^{-1}e_i,\qquad S(e_i)=-f_ik_i\qquad
S(k_i^{\pm 1})=k_i^{\pm 1},\qquad i=1,2.
\end{array}
\end{equation}
The ${\bf Z}_2$-grading of the generators are $[e_2]=[f_2]=1$ and zero
otherwise. The multiplication rule for the tensor product is
${\bf Z}_2$-graded and is defined for the elements $a,b,c,d$ of
$U_t(sl(2/1))$ by
\begin{equation}
(a\otimes b)(c\otimes d)=(-1)^{[b][c]}(ac\otimes bd).
\end{equation}

With the help of the symmetry operators Eq.(26), it is possible to give
the following construction for the generators $e_{1}$, $e_{2}$, $f_{1}$,
$f_{2}$, $k_{1}$ and $k_{2}$
\begin{equation}
\begin{array}{c}
e_{1}={T_{(m_{1},m_{2})}+T_{(-m_{1},m_{2})}
\over t-t^{-1}}\otimes {\bf 1},\qquad
f_{1}=-i{T_{(m_{1},-m_{2})}-T_{(-m_{1},-m_{2})}
\over t-t^{-1}}\otimes {\bf 1},\\
k_{1}=-iT_{(-2m_{1},0)}\otimes {\bf 1},\qquad
k_{1}^{-1}=iT_{(2m_{1},0)}\otimes {\bf 1},\\
k_{2}=-iT_{(-2m_{1},0)}\otimes\pmatrix{t^{-2}&0\cr
       0&t^{2}\cr},\qquad
k_{2}^{-1}=iT_{(2m_{1},0)}\otimes\pmatrix{t^{2}
       &0\cr
       0&t^{-2}\cr},\\
e_{2}={T_{(m_{1},-m_{2})}-T_{(-m_{1},-m_{2})}\over
{(t-t^{-1}})^{1\over 2}}\otimes f,\qquad
f_{2}={T_{(m_{1},m_{2})}+T_{(-m_{1},m_{2})}\over
{(t-t^{-1}})^{1\over 2}}\otimes f^{+}.\\
\end{array}
\end{equation}
It turns out that the above generators satisfy the algebraic relations
characterizing the quantum superalgebra $U_t(sl(2/1))$ as shown by Eq.(30)
where the $t$-deformed parameter is given by
\begin{equation}
t=e^{m_{1}m_{2}}.
\end{equation}
This is a way to prove that we can realize the $U_t(sl(2/1))$ by using
the supersymmetry operators stated above.

\section{Conclusion}
In this paper, we have shown how the graded $q$-pseudo-diffrential 
operators lead to obtain a supersymmetric extension of the sine algebra and 
the quantum algebra $U_t(sl(2))$. Otherwise, we have realized the 
the supersymmetric sine algebra and the quantum superalgebra
$U_t(sl(2/1))$ with the help of $q$-operators graded with a 
fermionic algebra.

\vspace{5mm}
\noindent

\section*{Acknowledgement}
A. Jellal wishes to thank Prof. S. Randjbar-Daemi, Head of High Energy 
Section of the Abdus Salam International Centre for Theoretical Physics 
(AS-ICTP), for the kind hospitality at his section. He would like also
to express his gratitute to Prof. E.H. Saidi for his encouragment. 
The authors are grateful to Prof. G. Thompson for reading the manuscript
and A. Jellal acknowledges him for his kind encouragment. A. Jellal is 
thankful to Prof. C.K. Zachos for drawing his attention to reference 
{\cite{11}}.

\end{document}